\documentclass[preprint,12pt]{elsarticle}

\usepackage{amssymb}
\usepackage[pdftex]{pict2e}  

\newcommand{\pend}{\hspace*{\fill} $\Box$}
\newtheorem{lemma}{Lemma}[section]
\newtheorem{theorem}[lemma]{Theorem}

\newtheorem{proposition}[lemma]{Proposition}

\journal{}

\begin{document}

\begin{frontmatter}



\title{A first efficient algorithm for enumerating all the extreme points of a bisubmodular polyhedron}


\author[Matsui]{Yasuko Matsui}
\author[Naitoh]{Takeshi Naitoh}
\author[Zhan]{Ping Zhan$^*$}

\affiliation[Matsui]{organization={Department of Mathematical Sciences School of Science, Tokai University},
            addressline={4-1-1, Kita-kaname}, 
            city={Hiratsuka},
            postcode={259-1294}, 
            state={Kanagawa},
            country={Japan}}
            
\affiliation[Naitoh]{organization={Faculty of Economics, Shiga University},
            addressline={1-1-1 Banba}, 
            city={Hikone},
            postcode={522-8522}, 
            state={Shiga},
            country={Japan}}
            
\affiliation[Zhan]{organization={Department of Communication and Business, Edogawa University},
            addressline={474 Komagi}, 
            city={Nagareyama},
            postcode={270-0198}, 
            state={Chiba},
            country={Japan}}           

\begin{abstract}
Efficiently enumerating all the extreme points of a polytope identified by a system of linear inequalities is a well-known challenge issue.
We consider a special case and present an algorithm that enumerates all the extreme points of a bisubmodular polyhedron in $\mathcal{O}(n^4|V|)$ time and $\mathcal{O}(n^2)$ space complexity, 
where $ n$ is the dimension of underlying space and $V$ is the set of  outputs.
We use the reverse search and signed poset linked to extreme points to avoid the redundant search. 
Our algorithm is a generalization of enumerating all the extreme points of a base polyhedron which comprises some combinatorial enumeration problems.
\end{abstract}



\begin{keyword}
 Submodular and bisubmodular functions \sep bidirected graphs \sep signed poset \sep Hasse diagram \sep reverse search

\end{keyword}

\end{frontmatter}


\section{Introduction}
\label{}
\noindent
\emph{\bf Motivation}: 
Enumeration for solutions is a well-studied problem \cite{af,kw2022,zhan,ziegler}, and can be found in the problems of graphs \cite{af,makino2004}. 
Its applications can be observed in several fields, e.g., more routes for evacuation \cite{yamanaka2018}, efficient and fair assignments of resources \cite{Ball2022}. 

It is a challenge to enumerate polynomially all the extreme points of a polyhedron defined by a system of linear inequalities \cite{klee, ziegler}. 
In this article, we treat a special case, i.e., propose a polynomial algorithm that enumerates all the extreme points of a bisubmodular polyhedron defined in Subsection \ref{ssec:bi}. 
\medskip

\noindent
\emph{\bf Related work}. The following two types of research are linked to our work:
\begin{itemize}
\vspace{-2mm}
\item[-] Reverse search for enumeration \cite{af} proposed by Avis and Fukuda has been employed in various enumeration algorithms. 
We adapt the reverse search in our algorithm.
\vspace{-2mm}
\item[-] Ando and Fujishige \cite{af1, af4} characterized the adjacency of the extreme points of a bisubmodular polyhedron by the associated signed poset. 
This property is crucial in avoiding redundant searches. 
\end{itemize}
\vspace{-2mm}
\medskip

For enumeration problems, if the complexity of an algorithm can be expressed as $p(x)l(y)$, where $p(x)$ is a polynomial function of input size $x$, and $l(y)$ is a linear function of the output size $y$, we call it {\it polynomial enumeration algorithm}. 
\medskip

\noindent
\emph {\bf The main contribution}: 
We define a local search function on signed posets associated with the extreme points, which makes a \emph{reasonable enumeration} \cite[Chapter 1]{KV2018} feasible. 
Our {\sf Algorithm 2} that enumerates all the extreme points of a bisubmodular polyhedron by adapting the reverse search {\sf Algorithm 1 }proposed by Avis and Fukuda. 
We show that {\sf Algorithm 2} runs $\mathcal{O}(n^4|V|)$ time and $\mathcal{O}(n^2)$ space complexity, and is $\mathcal{O}(n^6)$ time delay.
{\sf Algorithm 2} is a generalization of the enumeration algorithm on a base polyhedron \cite{zhan}.
\medskip

\noindent
\emph{\bf Possible applications}: Although bisubmodular polyhedron is a special case of a polyhedron identified by a system of linear inequalities.
The bisubmodular function is a generalization of the submodular  function which includes various combinatorial problems \cite{af, fuji2014}.    

Since inequalities on the entropy function are identical to the submodular condition, bisubmodular functions as examples of deep multivariate submodular functions also are applied to machine learning and data science \cite{bb2017}.
Other applications linked to bisubmodular functions, refer to the recent paper \cite{af1}.

Our algorithm is evidently applied to minimizing concave, or maximizing convex functions on a bisubmodular polyhedron. 
With recent advancements of submodularity in machine learning and artificial intelligence \cite{b2022} and the further generalization of bisubmodular functions \cite{af1}, our algorithm presents further possibilities of applications in these fields.
\medskip

\noindent
\emph{\bf Outline}:
The remainder of this article is structured as follows. 
In Section \ref{sec:def}, we introduce bisubmodular functions and polyhedra,
the adjacency of vertices characterized by the associated signed poset. 
The reverse search is also introduced later. 
Section \ref{sec:alg} presents our main result, a polynomial algorithm that enumerates all the extreme points of a bisubmodular polyhedron. 
Concluding remarks are given in Section \ref{sec:conclude}.
\section{Definitions and preliminaries} \label{sec:def}
\subsection{Bisubmodular polyhedra and the extreme point theorem} \label{ssec:bi}
Let $N=\{1,2,\cdots,n\}$ be a ground set.
Denote by $3^N$ the set of all ordered pairs of disjoint subsets of $N$, i.e., $3^N= \{(X,Y)\,|\,\, X,Y \subseteq N, X \cap Y = \emptyset \}$, and called such $(X,Y) \in 3^N$ a {\it signed subset\/} of $N$.  

Each $(X,Y)\in 3^N$ can be determined with its characteristic vector  \break $\chi_{(X,Y)} \in \{0,-1,1\}^N$ defined as follows: 

\begin{equation} 
\chi_{(X,Y)}(i)=
 \left\{ \begin{array}{rl} 1 & {\rm if} \; i \in X\\
   -1 & {\rm if} i \in Y\\
   0 &{\rm otherwise}
  \end{array}\right.\quad (i \in N).
\end{equation}

We define two binary operations, the {\it reduced union\/} $\sqcup$ and {\it intersection\/} $\sqcap $ on $3^N$ as
\begin{eqnarray}
 (X_1,Y_1)\sqcup (X_2,Y_2)&=&((X_1\cup X_2) \setminus (Y_1\cup Y_2),
(Y_1\cup Y_2) \setminus (X_1\cup X_2)),\label{eq:sqcup}\\
 (X_1,Y_1)\sqcap (X_2,Y_2)&=&(X_1\cap X_2,Y_1\cap Y_2)\label{eq:sqcap}
\end{eqnarray}
for each $(X_i,Y_i) \in 3^N$ ($i=1,2$). 
Examine \cite{af1} for some graph examples and details.  

In the following, for $i \in N$, we simplify $\{i\}$ as $i$ when it is clear from the context.
And for $X \subseteq N$ and $i \in N$ we write $X+i$ and $X-i$ rather than $X \cup \{i\}$ and $X \setminus \{i\}$, respectively. 

A biset function $f: 3^N \rightarrow \mathbb{R}$ is called {\it bisubmodular\/} if $f$ satisfies 
\begin{equation}
f(X_1,Y_1)+ f(X_2,Y_2) \geq f((X_1,Y_1) \sqcup(X_2,Y_2))+ 
   f((X_1,Y_1) \sqcap(X_2,Y_2))
 \label{eqn:bisub_in}
\end{equation}
for each $(X_i,Y_i) \in 3^N$ ($i=1,2$). 

Moreover, $f$ is called \emph{strict bisubmodular} if 
\begin{equation}
f(X_1,Y_1)+ f(X_2,Y_2) > f((X_1,Y_1) \sqcup(X_2,Y_2))+ 
   f((X_1,Y_1) \sqcap(X_2,Y_2))
 \label{eqn:bisub_strict}
\end{equation}
holds for all $(X_i,Y_i) \in 3^N$ such that $X_i - X_j \ne \emptyset$, 
or $Y_i - Y_j \ne \emptyset$ $(i=1,2, j=3-i)$.\footnote{These conditions are to exclude trivial case, where (\ref{eqn:bisub_in}) apparently are satisfied since equalities $\{(X_1,Y_1)\sqcup (X_2,Y_2),(X_1,Y_1)\sqcap (X_2,Y_2)\} = \{(X_1,Y_1), (X_2,Y_2) \}$.} 
(A strict bisubmodular function is depicted in Figure  \ref{fig:examps_bi_s}, the example of Figure \ref{fig:examps_bi_n} is not strict one.) 

For a bisubmodular function with $f(\emptyset,\emptyset)=0$, a pair $(3^N,f)$ is called a {\it bisubmodular system} on $N$. 
The {\it bisubmodular polyhedron\/} P$_*(f)$ associated with $(3^N,f)$  is defined as 
\begin{equation}
 {\rm P}_*(f)= \{ x \,\,| \,\,x \in \mathbb{R}^N, \forall (X,Y) \in 
3^N: \,\,\, x(X,Y) \leq f(X,Y)\},  \label{eqn:p_bi}
\end{equation}
where for each $(X,Y) \in 3^N$ and $x \in \mathbb{R}^N$,  
$ x(X,Y)=x(X)-x(Y)$ and ${ x(X)=\sum_{i \in X} x(i)}$ with $x(\emptyset)=0$. 
We have ${\rm P}_*(f) \ne \emptyset $ and ${\rm P}_*(f)$ is bounded 
(see more details about bisubmodular functions and polyhedra in \cite{af1,af4,fuji2014}).  

One of the key properties of bisubmodular polyhedra is the efficiency in calculating extreme points. 

We define a partial order $\sqsubseteq$ on $3^N$ satisfying 
\begin{equation} 
(X_1,Y_1)\sqsubseteq(X_2,Y_2)\Longleftrightarrow X_1\subseteq X_2, Y_1\subseteq Y_2,\; {\rm for} \ (X_i,Y_i) \in 3^N, \ i=1,2, \label{eq:sq_def_c}
\end{equation}
additionally, we write $(X_1,Y_1) \sqsubset (X_2,Y_2)$ if $X_1 \cup Y_1 \subset X_2 \cup Y_2$. 

The following theorem indicates a characterization of the extreme points of a bisubmodular polyhedron. 

\begin{theorem}[Extreme point theorem \cite{cu, bc, ck}] 
\label{th:ext}
A vector $x\in \mathbb{R}^N $ is an extreme point of ${\rm P}_*(f) $ if and only if there exists a chain  \\  
\begin{equation} \label{eq:chain_1}
 (U_0, W_0) \sqsubset (U_1, W_1) \sqsubset  \cdots \sqsubset (U_n, W_n) \quad (U_i,W_i) \in 3^N \,\,\,(i=0,1,\ldots,n)
\end{equation}
such that 
\begin{equation} 
f(U_i,W_i)-f(U_{i-1},W_{i-1})= \left\{ 
 \begin{array}{l} 
  x(U_i \setminus U_{i-1}) \quad \; {\it if} \; U_{i-1} \subset U_i \,({\it and }   \; W_{i-1}=W_i) \\
  \hspace{-1mm} -x(W_i \setminus W_{i-1}) \;  {\it if}\, W_{i-1} \subset W_i 
    \; ({\it and } \; U_{i-1}=U_i)
 \end{array} \right. \label{eq:chain_2}
\end{equation} 
for each $i=1,2, \ldots ,n$, where $(U_0, W_0)=(\emptyset,\emptyset)$.  
\end{theorem} 

We also call the extreme point a vertex if there is no ambiguity in the sequel. 

A signed pair $(S,T) \in 3^N$ is called {\it orthant} if $S \cup T=N$.  
The nonempty face of ${\rm P}_*(f)$ associated with an orthant $(S,T)$ satisfying 
\begin{equation} 
{\rm B}_{(S,T)}(f)=\{x\,|\,x \in {\rm P}_*(f),\,x(S,T)=f(S,T)\}
\end{equation} 
is called the {\it base polyhedron of} $(3^N, f)$ {\it of orthant} $(S,T)$. 
(The thick lines of ${\rm P}_*(f)$ in Figure \ref{fig:examps_bi_s} are ${\rm B}_{(S,T)}(f)$.) 

From Extreme point theorem \ref{th:ext}, we can represent ${\rm P}_*(f)$ as the convex hull of ${\rm B}_{(S,T)}(f)$.
Specifically, we have 
\begin{equation} \label{eq:orth}
{\rm P}_*(f)={\rm Conv}\{({\rm B}_{(S,T)}(f))\,| \, (S,T) 
\in 3^N, S \cup T=N\}. 
\end{equation}

\begin{proposition}
The number of extreme points of {\rm P}$_*(f)$ is at most $2^n n!$, the bound is tight. 
\medskip\\
{\rm (Proof)
For each orthant $(S,T)$, ${\rm B}_{(S,T)}(f)$ possesses the same combinatorial structure as a permutohedron (or the base polyhedron of a polymatroid \cite{f}), the number of its vertices is bounded by $n!$.
There are at most $2^n$ orthants $(S,T) \in 3^N$. 
From (\ref{eq:orth}), the number of vertices of {\rm P}$_*(f)$ is bounded  by $2^n n!$. 

The bound $2^n n!$ being tight can be verified by the following strict bisubmodular function 
 $f': 3^N \rightarrow \mathbb{R}$  
\begin{equation} \label{eq:bisub_strict_ex}
  f'(X,Y)  = \sum_{i=1}^{|X \cup Y|}(n+1 -i)
\end{equation}
with $f'(\emptyset,\emptyset) = 0$, where $\vert \cdot \vert $ means cardinality. 
\pend }
\end{proposition}
\subsection{Signed posets and the adjacency of extreme points} \label{ssec:signed}
A \emph{bidirected graph\/} is a $G=(V_G,A;\partial)$ with {\it boundary operator\/} $\partial\!:A\rightarrow \mathbb{Z}^{V_G}$ 
such that for each arc $a\in A$ there exists two vertices $i,j \in V_G$ conforming to one of the following three types: \par 
\smallskip 
 $\partial a = i-j$ \quad \,\, (arc $a$ has a tail at $i$ and a head at $j$),\par 
$\partial a = i+j$ \quad \,\, (arc $a$ has two tails, one at $i$, and the other at $j$), \par 
$\partial a = -i-j$ \quad (arc $a$ has two heads, one at $i$, and the other at $j$).  \par 
\smallskip
\noindent
If $i=j$, we call arc $a$ a {\it selfloop}. 
Any redundant selfloop $a$ of type $i-i$ is prohibited. 

A bidirected graph $(V_G,A;\partial)$ is called a 
{\it signed poset} if the following (i)$\sim $ (iii) are conformed to \cite{af1, af4, rein}.
\vspace{-1mm}
\begin{itemize}
\item[(i)] No two arcs $a, a' \in A$ with $\partial a=- \partial a'$ (acyclic) exist.
\vspace{-2.5mm}
\item[(ii)] For any two arcs $a,a' \in A$ that are not both selfloops 
and are oppositely incident to a typical vertex, there exists an arc $a'' \in A$ such that $\partial a'' = \partial a + \partial a'$ 
(transitivity).
\vspace{-2.5mm}
\item[(iii)] For any two selfloops $a, a'\in A$ incident to different vertices there exists an arc $a'' \in A$ such that $2 \partial a''= \partial a+\partial a'$ (transitivity).
\end{itemize}
\vspace{-1mm}

For a signed poset ${\cal P}=(V_G,A; \partial) $, we call $(X,Y) \in 3^{V_G}$ an {\it ideal} of ${\cal P}$ if  
\begin{equation}
 \langle \partial a, \chi_{(X,Y)} \rangle \leq 0 \ \qquad \forall a \in A, 
\label{eqn:ideal}
\end{equation}
where $\langle \cdot, \cdot \rangle$ denotes the ordinary inner product, and $\partial a$ is a mapping from $A$ to a free module with a base $V_G$. 
Denote by ${\cal I}({\cal P})$ the collection of all the ideals of the signed poset ${\cal P}$ \cite{Bir, rein}. 

For a bidirected graph $G=(V_G,A;\partial)$, if we add some arcs such that the new graph denoted by ${\tilde G}$ satisfying the transitivity properties, we call ${\tilde G}$ the {\it transitive closure} of $G$. 
Conversely, for an acyclic bidirected graph $G$, if ${\cal H}$ is the minimum bidirected graph satisfying ${\tilde {\cal H}}={\tilde G}$, we call ${\cal H}$ the {\it Hasse Diagram} of $G$, denoted by ${\cal H}(G)$.  
For each signed poset, its Hasse diagram is unique \cite{rein}. 

For $ x \in {\rm P}_*(f)$ we define ${\cal F}(x) \subseteq 3^N$ by
\begin{equation} \label{eq:f_x}
  {\cal F}(x)=\{(X,Y)\,\,| \,\,(X,Y) \in 3^N,\,\,x(X,Y)=f(X,Y)\}.\footnote{By our notation definitions, bisubmodular function $f$ is defined on $N$, not $V_G$.}  
\end{equation}
Then, we have the following result \cite{bc}: \par
\medskip 
 ($\dagger$) \qquad ${\cal F}(x)$ {is closed under} $\sqcap$ {and}  $\sqcup$ for $x \in {\rm P}_*(f) $. \par
\medskip 
We require more notations. 
For each $x \in {\rm P}_*(f)$ and each $ i \in N$, if we have 
\begin{equation} 
 \forall \alpha > 0: \,\,\, x+ \alpha \chi_i \notin {\rm P}_*(f),
\end{equation}
we say $x$ is {\it positively saturated at i}, where $\chi_i$ is a characteristic vector in $\{0,1\}^N$. 
Comparably, we say $x$ is {\it negatively saturated at i} if 
 \begin{equation} 
 \forall \alpha > 0: \,\,\, x- \alpha \chi_i \notin {\rm P}_*(f).
\end{equation}
Denote by sat$^{(+)}(x)$ (resp. sat$^{(-)}(x)) \subseteq N$ the positively (resp. negatively) saturated at $x \in {\rm P}_*(f)$, 
call sat$^{(+)}$ and sat$^{(-)}$ the {\it signed saturation function}. 

For each $i \in$sat$^{(+)}(x)$, define 
\begin{equation} \label{eq:dep_i}
 {\rm dep}(x,+i)(\equiv({\rm dep}(x,+i)^+,{\rm dep}(x,+i)^-))
   =\sqcap \{(X,Y)\,\,|\,\,i \in X,\,(X,Y) \in {\cal F}(x)\},
\end{equation}
and for $i\in$sat$^{(-)}(x)$, define 
\begin{equation} \label{eq:dep_-i}
 {\rm dep}(x,-i)(\equiv({\rm dep}(x,-i)^+,{\rm dep}(x,-i)^-))
  =\sqcap \{(X,Y)\,\,|\,\,i \in Y,\,(X,Y) \in {\cal F}(x)\}.
\end{equation}
We call dep the {\it signed dependence function}. 
(Refer to \cite{af4, f} for more details.) 

We present a signed poset associated with an extreme point $x \in {\rm P}_*(f)$.  

A bidirected graph $G({\cal F}(x))=(N,A(x);\partial)$ associated with $x \in {\rm P}_*(f)$ is defined as: \par
\smallskip 
\noindent
(1) For each $i \in N$, \par
 \hspace{-3mm} (1a) there is a selfloop $a$ at $i$ with $\partial a =2i$ 
	if and only if $i \in N-{\rm sat}^{(+)}(x),$ \par
  \hspace{-3mm} (1b) there is a selfloop $a$ at $i$ with $\partial a =-2i$ 
	if and only if $i \in N-{\rm sat}^{(-)}(x).$ \\
(2) For each different $i, j \in N$, \par 
\hspace{-2mm}(2a) there is an arc $a$ with $\partial a = i-j$ if and only if 
 $j \in {\rm dep}(x,+i)^+$ or $ i \in {\rm dep}(x,-j)^-,$\par
\hspace{-3mm} (2b) there is an arc $a$ with $\partial a = i+j$ if and only if 
 $j \in {\rm dep}(x,+i)^-$ or $ i \in {\rm dep}(x,+j)^-,$\par
\hspace{-3mm} (2c) there is an arc $a$ with $\partial a = -i-j$ if and only if 
 $j \in {\rm dep}(x,-i)^+$ or $ i \in {\rm dep}(x,-j)^+.$\par
\smallskip 
It can be shown \cite{af1,af4} that $G({\cal F}(x))$ is a signed poset, i.e., satisfying acyclic property, and ${\cal I}(G({\cal F}(x)))={\cal F}(x)$. 
We write $G({\cal F}(x))$,  
${\cal I}(G({\cal F}(x)))$ and ${\cal H}(G({\cal F}(x)))$ 
as $G(x$), ${\cal I}(x)$ and ${\cal H}(x)$ respectively for simplicity in the sequel. 
For the extreme point $x \in$ P$_*(f)$, its $G(x)$ and the dep function can be calculated in O$(n^2$) \cite{af1,af4}. 

An example that satisfies the strict bisubmodular inequality (\ref{eqn:bisub_strict}) is depicted in Figure \ref{fig:examps_bi_s}. 
The adjacent vertices $x_1$ and $x_2$ of the bisubmodular polyhedron are calculated by a signed chain.
The diagram of $G(x_i)$, ${\cal I}(x_i)$, and ${\cal H}(x_i)$ ($i=1,2)$ are constructed based on the signed saturated and the dep subsets, respectively. 
The paths in ${\cal I}(x_1)$ and ${\cal I}(x_2)$ are drawn from the bottoms. 
(Notation $\tilde G(x_i)$ will be defined in Theorem \ref{th:adjac}.)

The case of non-strict bisubmodular functions is given in Figure \ref{fig:examps_bi_n}. 
Refer to \cite{msz_2023} for more bisubmodualr polyhedron examples.

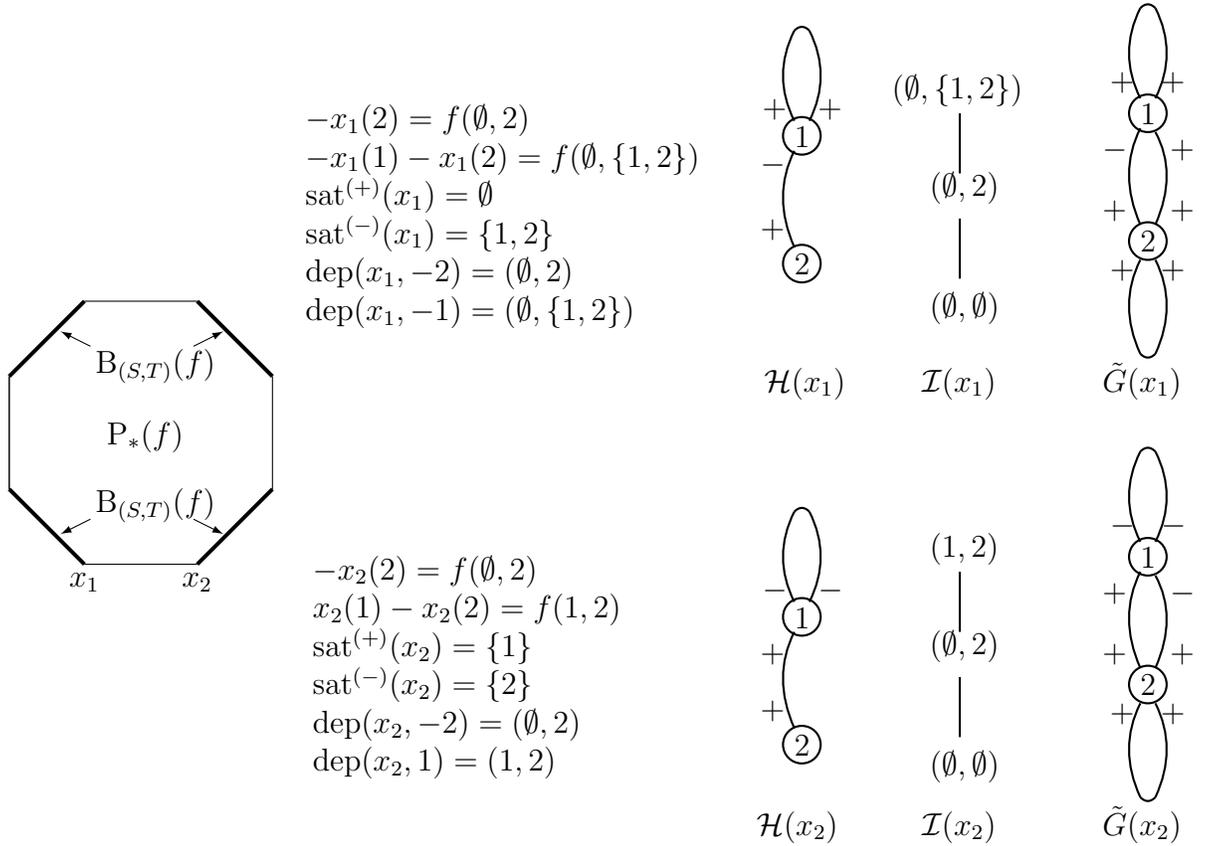
\begin{figure}[h]

\setlength{\unitlength}{1mm}
\begin{picture}(163,115)(0,0)

\put(12,40){\line(1,0){15}} \put(37,50){\line(0,1){15}}
\put(2,50){\line(0,1){15}} \put(2,50){\line(0,1){15}}
\put(12,75){\line(1,0){15}}  \put(15,56){P$_*(f)$}
\put(10,37){$x_1$} \put(25,37){$x_2$} 

\put(13.5,65.5){B$_{(S,T)}(f)$}
\put(12.8,69){\vector(-2,1){4}} \put(26.5,69){\vector(2,1){4}}
\put(13.5,47){B$_{(S,T)}(f)$}  
\put(12.8,46){\vector(-2,-1){4}} \put(26.5,46){\vector(2,-1){4}}

\linethickness{1.5pt}
\put(27,40){\line(1,1){10}}
\put(12,40){\line(-1,1){10}}
\put(2,65){\line(1,1){10}}\put(37,65){\line(-1,1){10}}

\linethickness{0.8pt}
\put(40,98){ \begin{minipage}[t]{53mm}
$-x_1(2) = f(\emptyset,2)$ \\
$-x_1(1)-x_1(2) = f(\emptyset,\{1,2\})$\\
sat$^{(+)}(x_1)=\emptyset$\\
sat$^{(-)}(x_1)=\{1,2\}$ \\
dep$(x_1,-2)=(\emptyset,2)$ \\
dep$(x_1,-1)=(\emptyset,\{1,2\})$ \\
\end{minipage}}

\put(148,78){ \begin{minipage}[t]{24mm}
\begin{picture}(14,30)(0,0)
\qbezier(3,24)(0,30)(3,36) \qbezier(5,24)(8,30)(5,36)
\qbezier(3,36)(4,37)(5,36) 
\put(-1,25){$+\;\;\; +$}
\put(4,22){\circle{5}}\put(3,20.5){$1$}
\put(-2,16){$-$}  \put(-2,8){$+$}
\put(7,16){$+$}  \put(7,8){$+$}
\put(4,5){\circle{5}} \put(3,3.6){$2$}
\put(-1,0){$+ \;\;\; +$}
\qbezier(3,-10)(0,-3.5)(3,2.5) \qbezier(5,-10)(8,-3.5)(5,2.5)
\qbezier(3,-10)(4,-11)(5,-10) 
\qbezier(3,20)(0,14)(3,7.2) 
\qbezier(5,19.5)(8,14)(5,7.2) 
\put(-2,-15){$\tilde G(x_1)$}
\end{picture}
\end{minipage} }

\put(122,73){ \begin{minipage}[t]{24mm}
\begin{picture}(14,30)(0,0)
\put(1,0){$(\emptyset,\emptyset)$} \put(5,5){\line(0,1){8}}
\put(1,16){$(\emptyset,2)$} \put(5,19){\line(0,1){8}} 
\put(-4,29){$(\emptyset,\{1,2\})$}
\put(-0,-10){${\cal I}(x_1)$}
\end{picture}
\end{minipage} }

\put(102,75){ \begin{minipage}[t]{18mm}
\begin{picture}(20,30)(0,0)
\qbezier(3,24)(0,30)(3,36) \qbezier(5,24)(8,30)(5,36)
\qbezier(3,36)(4,37)(5,36) 
\put(-1.2,24.5){$+\quad +$}
\put(4,22){\circle{5}}\put(3,20.5){$1$}
\qbezier(3,20)(0,14)(3,7.2) 
\put(-1.5,17){$-$}  \put(-1.5,8.5){$+$}
\put(4,5){\circle{5}} \put(3,3.6){$2$}
\put(-1,-12){${\cal H}(x_1)$}
\end{picture}
\end{minipage} }
\put(41,38){ \begin{minipage}[t]{58mm}
$-x_2(2) = f(\emptyset,2)$ \\
$x_2(1)-x_2(2) = f(1,2)$\\
sat$^{(+)}(x_2)=\{1\}$\\
sat$^{(-)}(x_2)=\{2\}$ \\
dep$(x_2,-2)=(\emptyset,2)$ \\
dep$(x_2,1)=(1,2)$ \\
\end{minipage} }

\put(148,19){ \begin{minipage}[t]{23mm}
\begin{picture}(14,30)(0,0)
\qbezier(3,24)(0,30)(3,36) \qbezier(5,24)(8,30)(5,36)
\qbezier(3,36)(4,37)(5,36) 
\put(-1,25){$-\;\;\; -$}
\put(4,22){\circle{5}}\put(3,20.5){$1$}
\put(-2,16){$+$}  \put(-2,8){$+$}
\put(7,16){$-$}  \put(7,8){$+$}
\put(4,5){\circle{5}} \put(3,3.6){$2$}
\put(-1,0){$+ \;\;\; +$}
\qbezier(3,-10)(0,-3.5)(3,2.5) \qbezier(5,-10)(8,-3.5)(5,2.5)
\qbezier(3,-10)(4,-11)(5,-10) 
\qbezier(3,20)(0,14)(3,7.2) 
\qbezier(5,19.5)(8,14)(5,7.2) 
\put(-2,-15){$\tilde G(x_2)$}
\end{picture}
\end{minipage} }

\put(122,12){ \begin{minipage}[t]{23mm}
\begin{picture}(14,30)(0,0)
\put(1,0){$(\emptyset, \emptyset)$} \put(5,5){\line(0,2){8}}
\put(1,16){$(\emptyset,2)$} \put(5,19){\line(0,1){8}} \put(1,29){$(1,2)$}
\put(-0,-8){${\cal I}(x_2)$}
\end{picture}
\end{minipage} }

\put(102,11){ \begin{minipage}[t]{18mm}
\begin{picture}(20,30)(0,0)
\qbezier(3,24)(0,30)(3,36) \qbezier(5,24)(8,30)(5,36)
\qbezier(3,36)(4,37)(5,36) 
\put(-1.2,24.5){$-\quad -$}
\put(4,22){\circle{5}}\put(3,20.5){$1$}
\qbezier(3,20)(0,14)(3,7.2) 
\put(-1.5,16){$+$}  \put(-1.5,8.5){$+$}
\put(4,5){\circle{5}} \put(3,3.6){$2$}
\put(-2,-7){${\cal H}(x_2)$}
\end{picture}
\end{minipage} }

\end{picture}
\caption{An example of the strict bisubmodular function. (The flow of the computations: $x_i \to {\rm sat}(x_i) \to {\rm dep}(x_i) \to (G(x_i)) \ {\cal H}(x_i ) \to {\cal I}(x_i) \to \tilde G(x_i)$, $i=1,2$). }
\label{fig:examps_bi_s}
\end{figure}
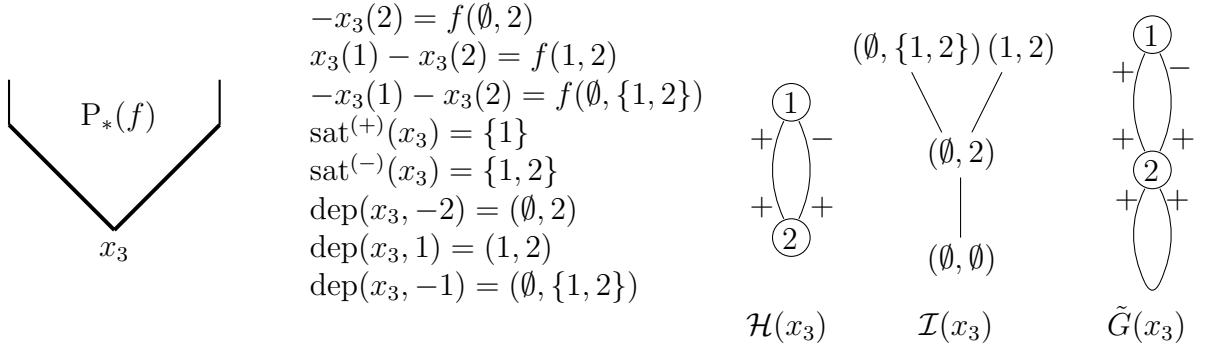
\begin{figure}[h]

\setlength{\unitlength}{1mm}
\begin{picture}(170,50)(0,0)

\put(10,10){ \begin{minipage}[t]{23mm}
\begin{picture}(24,30)(0,0)
\linethickness{1.5pt}
\put(0.5,22){P$_*(f)$}
\put(5,8){\line(-1,1){14}} \put(5,8){\line(1,1){14}} 
\linethickness{0.8pt}
\put(-9,22){\line(0,1){6}} \put(19,22){\line(0,1){6}} 
\put(3,4.5){$x_3$}
\end{picture}
\end{minipage} }

\put(41,45){ \begin{minipage}[t]{58mm}
$-x_3(2) = f(\emptyset,2)$ \\
$x_3(1)-x_3(2) = f(1,2)$\\
$-x_3(1)-x_3(2) = f(\emptyset,\{1,2\})$\\
sat$^{(+)}(x_3)=\{1\}$\\
sat$^{(-)}(x_3)=\{1,2\}$ \\
dep$(x_3,-2)=(\emptyset,2)$ \\
dep$(x_3,1)=(1,2)$ \\
dep$(x_3,-1)=(\emptyset,\{1,2\}) $\\
\end{minipage} }

\put(149,11){ \begin{minipage}[t]{23mm}
\begin{picture}(14,30)(0,0)
\put(4,33){\circle{5}}\put(2.8,31.5){1}
\qbezier(3,30.5)(0,24)(3,17.5) \qbezier(5,30.5)(8,24)(5,17.5)
\put(-1.5,27){$+$}  \put(-1.5,18){$+$}
 \put(6,27.5){$-$}  \put(6.5,18){$+$}
\put(4,15){\circle{5}} \put(2.8,13.5){2}
\qbezier(3,0)(0,6.5)(3,12.5) \qbezier(5,0)(8,6.5)(5,12.5)
\qbezier(3,0)(4,-2)(5,0) 
\put(-1,10){$+ \;\;\; +$}
\put(-2,-7){$\tilde G(x_3)$}
\end{picture}
\end{minipage} }

\put(122,13){ \begin{minipage}[t]{23mm}
\begin{picture}(14,30)(0,0)
\put(1,0){$(\emptyset, \emptyset)$} \put(5.5,4){\line(0,1){8}} \put(1,14){$(\emptyset,2)$} 
\put(3,18){\line(-1,2){4}} \put(-9,28){$(\emptyset,\{1,2\})$}
\put(7,18){\line(1,2){4}} \put(9,28){$(1, 2)$}
\put(-0,-9){${\cal I}(x_3)$}
\end{picture}
\end{minipage} }

\put(101,12){ \begin{minipage}[t]{18mm}
\begin{picture}(20,30)(0,0)
\put(4,23){\circle{5}}\put(2.9,21.6){1}
\qbezier(3,7.5)(0,14)(3,20.5)
 \put(-1.5,17){$+$}  \put(-1.5,8){$+$}
 \qbezier(5,7.5)(8,14)(5,20.5)
\put(6.5,17){$-$}  \put(6.5,8){$+$}
\put(4,5){\circle{5}} \put(2.8,3.5){2}
\put(-2,-8){${\cal H}(x_3)$}
\end{picture}
\end{minipage} }

\end{picture}\par \smallskip \par
\caption{An example of the non-strict bisubmodular function.}

\label{fig:examps_bi_n}
\end{figure}

We are ready to introduce an important theorem that characterizes the adjacency of the extreme points by associated signed posets.
  
Let $x$ be a vertex of {\rm P}$_*(f)$. 
For each arc $a$ of ${\cal H}(x)$, let $G^\prime(x,a)$ be the bidirected graph acquired by adding to $G(x)$ the arc ${\bar a}$ with the boundary $ \partial {\bar a}= -\partial a$. 
We have:
\begin{theorem}[\cite{af4}] \label{th:adjac}
Two distinct vertices $x$ and $x'$ of {\rm P}$_*(f)$ are adjacent if and only if there exist arcs $a$ of ${\cal H}(x)$ and $a'$ of ${\cal H}(x')\,\,\, (\partial a = -\partial a')$ 
such that ${\tilde G}^\prime (x,a)={\tilde G}^\prime (x',a')$, where ${\tilde G}^\prime (x,a)$ and ${\tilde G}^\prime (x',a')$ are the transitive closures of $G^\prime (x,a)$ and 
$G^\prime (x',a')$, respectively. 
\end{theorem}

For $x \in {\rm P}_*(f)$ and $i,j \in N$,  
$c(x, i, j)$ and $c(x, i, -j)$ are the {\it capacity functions} defined as  
\begin{equation} \label{eq:cap1}
  c(x,i,j)
  =\max\{\alpha \mid \alpha \in \mathbb{R}_{+}, 
  x+\alpha(\chi_{i}+ \chi_{j}) \in {\rm P}_*(f)\}, 
\end{equation}
and 
\begin{equation}\label{eq:cap2}
  c(x,i,-j)
  =\max\{\alpha \mid \alpha \in \mathbb{R}_{+},   
  x+\alpha(\chi_{i}- \chi_{j}) \in {\rm P}_*(f)\},  
\end{equation}
where $\chi_i$ and $\chi_j$ are the characteristic vectors. 
The other signed cases are addressed similarly.

In the following, we identify an arc with its partial if there is no ambiguity.
\subsection{Reverse search for enumeration} \label{sec:rs}
Our algorithm employs the scheme of the {\it reverse search for efficient enumeration} suggested by Avis and Fukuda \cite{af}. 
 
Let $G_R=(V_R,E_R)$ be a search graph. 
For a subset $S \subset V_R$, a triple $(G_R,S,g)$ is called a \emph{local search} if $g$: $V_R \setminus  S \to S$ satisfying $\{v,g(v)\} \in E_R $.  
Function $g$ is called the \emph{local search function}.  

Moreover, let $T_R=(V_R,A_R(g))$ be a directed spanning tree comprising a unique sink $s$ and $A_R(g)=\{(v,g(v)) \mid v \in V_R \setminus S\}$, the \emph{(abstract) reverse search} $(G_R, S,g)$ is: from $s\in S$, traverse $T_R(s)$, and output all the vertices of $G_R$.

For the linear programming, $V_R$ is the set of feasible bases, $T_R$ is induced by the pivot operation, $g$ is the (steepest) pivot, and $S$ is the (unique) optimal solution $s$.

Let {\it Adj} be a finite adjacent list, and ${\it Adj}(v)$ be the vertex adjacent to the vertex $v$ associated with the directed spanning tree $T_R$. 
The enumeration of the reverse search for a triple $({\it Adj},s,g )$ of $G_R=(V_R,E_R)$ is presented as the following {\sf Algorithm 1}.

\noindent
---------------------------------------------------------------------------------------------------\\
{\sf Algorithm 1: Reverse search for enumeration} \cite{af} \\
---------------------------------------------------------------------------------------------------\\
{\bf Input:} A triple $({\it Adj}, s, g)$ of $G_R=(V_R,E_R)$.\smallskip \\
{\bf Output:} All vertices $V_R$ of $G_R$.\smallskip \\
{\bf Initialize:} Put $v=s$. \smallskip \\
{\bf Reverse search:} Put $next= Adj(v)$.\footnote{We assume that $Adj$ is the two-oriented list of the data structure, each operation of $next= Adj(v)$ gets a new ``next'' element on the list.} If $next \ne \emptyset$, go to Reverse traverse. \par 
Otherwise ($next = \emptyset$), go to Forward traverse. \par
\noindent
{\bf Reverse traverse:} If $g(next)=v$, put $v=next$.  
(Otherwise, keep $v$.) \par  
Go to Reverse search. \smallskip \par
\noindent
{\bf Forward traverse:} If $v = s$, stop. \par 
Otherwise, put $u=v, v=g(v)$ and $Adj(v)=u.$ 
Go to Reverse search. \\
--------------------------------------------------------------------------------------------------
\section{A polynomial enumeration algorithm} \label{sec:alg}
To avoid the redundant search and give a polynomial enumeration algorithm, it is critical to define properly a local search function $g$ introduced in Subsection \ref{sec:rs} on ${\rm P}_*(f)$.

Without loss of generality, let $w \in \mathbb{R}^n $ satisfying $w(n)\gg \cdots \gg w(1)>0$. 
Consider 
\begin{equation}
  \max_{x \in {\rm P}_*(f)} \sum_{i=1}^n w(i)x(i). \label{eq:max_g}
\end{equation} 
The optimal solution of (\ref{eq:max_g}), i.e., $s$ in {\sf Algorithm 1}, is calculated by Extreme point theorem \ref{th:ext} as follows:     
\begin{eqnarray}
 & &  x^*(n)=f(n, \emptyset), \label{eq:x^*_1}  \\
 & &  x^*(i)=f(\{n,\dots,i\}, \emptyset)-f(\{n,\dots,i+1\}, 
\emptyset) \quad (i=n-1,\dots,1). \label{eq:x^*_2}
\end{eqnarray}

For each vertex $x \in {\rm P}_*(f)$, the local search function $g: V \to Adj$ based on the Theorem \ref{th:adjac} is defined as: 
\begin{equation}
  g(x)=x+ c(x, \pm i^*, j^*) ({j^*} \pm {i^*}), \label{eq:local_f}
\end{equation}
where $\pm \ i^*+ j^* = \partial a^* \in \mathcal{H}(x)$, and $j^*,i^*$ are the lexical maximums among arcs in ${\cal H}(x)$. 
Precisely, 
\begin{equation} \label{eq:j^*}
j^* = \max\{j \mid  \partial a = j \pm i \in {\cal H}(x) \; {\it for} \;  j \ge i \}, 
\end{equation}
\vspace{-10mm}
\begin{eqnarray} 
 \; \; \; i^* =  \left\{\begin{array}{l} \max\{ \ i \mid 
   {\it if}\;\partial a = j^* + i \;{\it in}\;(\ref{eq:j^*}) \}, \\
 \max\{\ i\mid {\it  if}\; \partial a = j^*- i \; {\it in} \; (\ref{eq:j^*})  \}.
\end{array}  \right.   \label{eq:i^*}
\end{eqnarray} 
In (\ref{eq:i^*}) (and (\ref{eq:j^*})), if $ \max\{j \mid  \partial a= j+i \}= \max\{j \mid \partial a= j -i \} $, we select $i^*$ of type $i+j^*$ by the lexical maximum.

We identify the local search function $g$ by the arc defined in (\ref{eq:local_f}) in its Hasse diagram if there is no ambiguity.  

The local search function $g$ can be described as follows. 
If arc $a$ in (\ref{eq:i^*}) is a type of $i^*+j^*$ (including the selfloop), we increase the values of the elements on the axes $i^*$ and $j^*$ by the traverse. 
Otherwise (the type of $j^*-i^*$ is chosen), we elevate the value of the element on the larger index and decrease the value on the smaller index. 

The local search function $g$ defined by (\ref{eq:local_f}) possesses a \emph{global} property as evidenced in Proposition \ref{th:valif2}.
\begin{proposition} \label{th:valif2}
For the extreme point $ x \in {\rm P}_*(f)$, there is no arc $a$ of type  $ \partial a = i + j $ ($i \le j$) or $\partial a=j-i$ ($i<j$) in Hasse diagram ${\cal H}(x)$ if and only if $x=x^*$. 
\medskip \\
{\rm (Proof) 
We first show Claim 1 and Claim 2.
\smallskip \\
{\bf Claim 1:} \emph{No arc of type $ i +j$ in ${\cal H}(x)$ if and only if no arc of type $ i +j$ in signed poset $G(x)$. }
\smallskip

No arc of type $i+j$ in $G(x)$ clearly means no arc of type $i+j$ in ${\cal H}(x)$. 
Conversely, if there exists no arc of type $i+j$ in ${\cal H}(x)$, it is impossible to create an arc of type $ i+j$ in the associated $G(x)$ 
by two transitivity operations (ii) and (iii) presented in Subsection \ref{ssec:signed}. 
This ends the proof of Claim 1.
 \smallskip \\
{\bf Claim 2:} \emph{No arc of type $i+j$ in ${\cal H}(x)$ if and only if $x \in {\rm B}_{(N, \emptyset)}(f)$.}
 \smallskip 
 
By Claim 1, we have that there exists no arc of type $ \partial a=i+j$ in $G(x)$ if no such type exists in ${\cal H}(x)$.  

In the case $i=j$, if no selfloop $2i$ exists in $G(x)$, we have sat$^{(+)}(x)=N$ (the definition (1a) of Subsection \ref{ssec:signed}).  

In the case $i\ne j$, by the definition (2b) of $G(x)$, we get
\begin{equation} \label{eq:claim_2}
 {\rm dep}(x,+i)^- = \emptyset \qquad \forall i \in N.  
\end{equation}
By taking a reduced union on (\ref{eq:claim_2}) for all $i \in N$, we obtain
 $${\displaystyle \sqcup_{i \in N} {\rm dep}(x,+i) = 
(\cup _{i \in N} {\rm dep}(x,+i)^+, \emptyset) =(N,\emptyset). }
$$
Then, we have $(N, \emptyset) \in {\cal F}(x)$ and $x \in {\rm} B_{(N, \emptyset)}(f)$ since ${\cal F}(x)$ is closed under the operation $\sqcup $ as shown in ($\dagger$). 
 
Conversely, suppose $x \in {\rm} B_{(N, \emptyset)}(f)$. 
It is clear that ${\rm sat} ^{(+)}(x)=N$. 
Then, no arc of type $2i$ ($i \in N$) exists in $G(x)$, and also does not exist in ${\cal H}(x)$ based on Claim 1. 
For type $i+j$ ($i \ne j$), we prove it by contradiction. 
If there exists type $ i+j$ in $G(x)$, then we have $i\in {\rm dep}(x,+j)^-$ (or $j\in {\rm dep }(x,+i)^-$), 
which contradicts the definition of dep functions since ${\rm dep }(x,+i)^-=\emptyset$ for each $i\in N$ after the reduced intersection with $(N,\emptyset)$. 
This ends the proof of Claim 2. 
\smallskip 

For the remainder of the proposition, the fact that there is no type $j-i$ ($i < j$) in ${\cal H}(x)$ for $x \in B_{(N,\emptyset)}(f)$ 
if and only if $x=x^*$, can be procured from the same discussion as the base polyhedron of a polymatroid \cite{zhan}. 
 \pend
}
\end{proposition}

Proposition \ref{th:valif2} shows the validity of the local search function $g$ defined in (\ref{eq:local_f}).

Next, we address the efficiency. 
Our goal is to skip the step of traversing to the adjacent vertices to verify $(v,g(v))\in A(g)$ based on the newly constructed Hasse diagrams. 
In other words, we want to know the local search function $g$ defined in (\ref{eq:local_f}) at the current vertex. 

Precisely, let ${\hat x}=g(x)$ be a vertex of ${\rm P}_*(f)$. 
Then, the arcs of types $j^* \pm i^* $ defined in (\ref{eq:local_f}) associated with $\mathcal{H}(x) $ are contained in $A^{\hat x}_R$ define as follows:
\begin{equation} \label{eq:A_R}
 A^{\hat x}_R= \{ a \in \mathcal{ H}({\hat x}) \mid \partial a =  \mp i -j \ {\it for }\ i \le j \}  
\end{equation}
i.e., reversing the sign $j^* \pm i^* $ of (\ref{eq:local_f}) based on the adjacent property revealed in Theorem \ref{th:adjac}.

We have $A^{\hat x}_R \ne \emptyset$ by Proposition \ref{th:valif2} if P$_*(f)$ is not a single point. 

We define a (signed) \emph{lexicographically order} $\prec_{\hat x}$ on $A^{\hat x}_R $ as follows.    
For $\partial a= \pm i - j$, $\partial a' =\pm i' - j $, and $\partial a''=\pm i-j^\prime$ in $A^{\hat x}_R$, 
we say $a\prec_{\hat x}a'\prec_{\hat x}a''$ if $j< j'$, $i < i'$ (recall $i,i'\le j,j'$),
and for arcs with the same $i,j$ and different types of the sign, $\partial a =i-j$ and $\partial a' =-i-j$, let $a \prec_{\hat x} a'$. 
\smallskip

\noindent
---------------------------------------------------------------------------------------------------\\
{\sf Algorithm 2: Enumerating all vertices of a bisubmodular polyhedron}\\
---------------------------------------------------------------------------------------------------\\
\noindent {\bf Input:} A bisubmodular function $f$: $3^N \rightarrow \mathbb{R}$.\smallskip \\
{\bf Output:} All the vertices of bisubmodular polyhedron P$_*(f)$.\smallskip \\
{\bf Initialize:} Compute vertex $x^*$ by (\ref{eq:x^*_1}) and (\ref{eq:x^*_2}). \par
Output $x^*$, put ${\hat x}=x^*$ and  call {Procedure 1}.
\smallskip \\
{\bf Reverse search:} Select the next arc from the arc list $A^{\hat x}_R$, go to Reverse traverse. \par  
Otherwise ($A^{\hat x}_R=\emptyset$, or all the arcs of $A^{\hat x}_R$ searched), go to Forward traverse. \smallskip \\
{\bf Reverse traverse:} Compute an adjacent vertex by 
\vspace{-2mm}
\begin{equation} \vspace{-2mm}
x={\hat x}+c({\hat x},\mp i, -j)(\mp \chi_i- \chi_j),  \label{eq:alg_2_g}
\end{equation}
\quad \; and call Procedure 1. 
If $g(x)={\hat x}$, output $x$, put ${\hat x} = x$. \par
Go to Reverse search. 
\smallskip \\ 
{\bf Forward traverse:} If ${\hat x} = x^*$, stop. \par 
Otherwise, compute $g({\hat x})$ by 
\vspace{-2mm}
\begin{equation} \vspace{-2mm}
x={\hat x}+c({\hat x},\pm i^*, j^*)(\pm \chi_{i^*}+ \chi_{j^*}), \label{eq:alg_2_g2}
\end{equation} 
\quad \; and call Procedure 1. \par 
Put ${\hat x}=x$. Go to Reverse search with $A^{\hat x}_R$ being pointed at $\pm i^*,j^*$.  
\medskip \\
{\bf Procedure 1, Construct Hasse diagram:} 
Compute $\mathcal{H}(x)$ and $A^x_R$. \\
---------------------------------------------------------------------------------------------------
\\
We apply the lexicographically \emph{ordered list} of $A^{\hat x}_R $ in the above Reverse search.
\bigskip

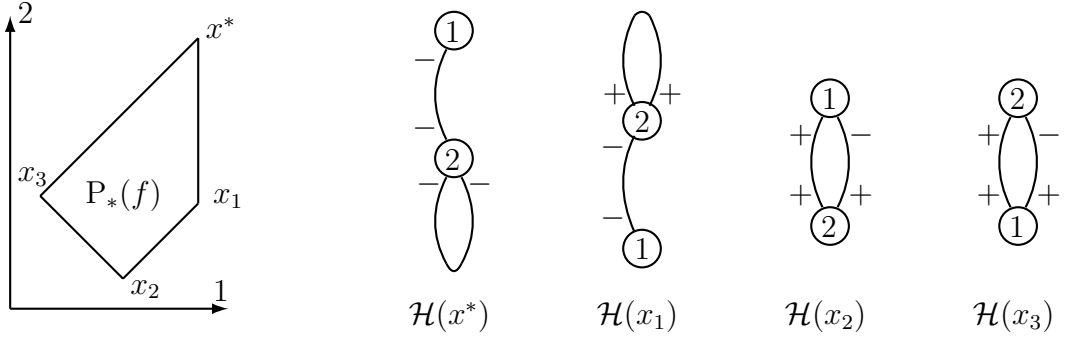
\begin{figure}[h]
\setlength{\unitlength}{1mm}
\begin{picture}(150,45)(0,0)

\linethickness{0.8pt}
\put(7,4){ \begin{minipage}[t]{40mm}
\begin{picture}(24,30)(0,0)
\put(4,15){\line(1,1){21}}
\put(25,36){\line(0,-1){22}}
\put(15,4){\line(-1,1){11}} \put(15,4){\line(1,1){10}}
\put(10,14){P$_*(f)$} 
\put(26,36){$x^*$} \put(27,14){$x_1$}
\put(16,2){$x_2$} \put(1,17){$x_3$} 
\put(0,0){\vector(1,0){29}} \put(27,1){1}
\put(0,0){\vector(0,1){39}} \put(1,38){2}
\end{picture}
\end{minipage} }

\put(62,9){ \begin{minipage}[t]{20mm}
\begin{picture}(20,30)(0,0)
\put(4,32){\circle{5}}\put(2.8,30.5){1}
\qbezier(3,29.5)(0,23.5)(3,17.5) 
\put(-1.5,27){$-$}  \put(-1.5,18){$-$}
\put(4,15){\circle{5}} \put(2.8,13.5){2}
\qbezier(3,1)(0,6.5)(3,12.5) \qbezier(5,1)(8,6.5)(5,12.5)
\qbezier(3,1)(4,-1)(5,1) 
\put(-1,10.5){$- \;\;\; -$}
\put(-2,-7){${\cal H}(x^*)$}
\end{picture}
\end{minipage} }

\put(87,7){ \begin{minipage}[t]{20mm}
\begin{picture}(20,30)(0,0)
\qbezier(3,24)(0,30)(3,36) \qbezier(5,24)(8,30)(5,36)
\qbezier(3,36)(4,37)(5,36) 
\put(-1.2,24.5){$+\quad +$}
\put(4,22){\circle{5}}\put(3,20.5){$2$}
\qbezier(3,20)(0,14)(3,7.2) 
\put(-1.5,17.5){$-$}  \put(-1.5,8){$-$}
\put(4,5){\circle{5}} \put(3,3.6){$1$}
\put(-2,-5){${\cal H}(x_1)$}
\end{picture}
\end{minipage} }

\put(112,10){ \begin{minipage}[t]{20mm}
\begin{picture}(20,30)(0,0)
\put(4,22){\circle{5}}\put(2.9,20.6){1}
\qbezier(3,7.5)(0,13.5)(3,19.5)
 \put(-1.5,16){$+$}  \put(-1.5,8){$+$}
 \qbezier(5,7.5)(8,13.5)(5,19.5)
\put(6.5,16){$-$}  \put(6.5,8){$+$}
\put(4,5){\circle{5}} \put(2.8,3.5){2}
\put(-2,-8){${\cal H}(x_2)$}
\end{picture}
\end{minipage} }

\put(137,10){ \begin{minipage}[t]{20mm}
\begin{picture}(20,30)(0,0)
\put(4,22){\circle{5}}\put(2.9,20.6){2}
\qbezier(3,7.5)(0,13.5)(3,19.5)
 \put(-1.5,16){$+$}  \put(-1.5,8){$+$}
 \qbezier(5,7.5)(8,13.5)(5,19.5)
\put(6.5,16){$-$}  \put(6.5,8){$+$}
\put(4,5){\circle{5}} \put(2.8,3.5){1}
\put(-2,-8){${\cal H}(x_3)$}
\end{picture}
\end{minipage} } \\
\end{picture} 
\caption{An example of the execution of {\sf Algorithm 2.}}
\label{fig:algorith}
\end{figure}

\noindent
{\bf Example 1:}
Consider a bisubmodular polyhedron P$_*(f)$ illustrated on the left of 
Figure \ref{fig:algorith}. 
The Hasse diagrams of all vertices are given on the right, respectively. 

Starting from ${\hat x}=x^*$, the Hasse diagram $\mathcal{H}(x^*)$ is built after its dep function (defined in Subsection \ref{ssec:signed}) is computed. 
Both arcs of $\mathcal{H}(x^*)$, ordered by $-1-2 \prec_{x^*} -2(2)$, are on the list of $A^{x^*}_R$. 

In Reverse search, arc $-2(2)$ is chosen according to the order $-1-2 \prec_{x^*} -2(2)$. 

In Reverse traverse, compute adjacent vertex $x_1$ using arc $-2(2)$. Then, compute $\mathcal{H}(x_1)$. 
Since $g(x_1)=x^*$, which means that function $g$ of (\ref{eq:local_f}) chooses arc $2(2)$ of $\mathcal{H}(x_1)$, we have $-(-2(2))=2(2)$ being the arc of the searching tree ($T_R$ in {\sf Algorithm 1}), output $x_1$.
Put ${\hat x}=x_1$.

In the next Reverse search, by similar arguments, we obtain $x_2$,  $\mathcal{H}(x_2), $ and $A^{x_2}_R $ from $\mathcal{H}(x_2)$. 
The Forward traverse $x_2 \to x_1$ is performed.    

For the rest of the execution, refer to Table \ref{tab:algorithm_t} for a summary. 

\begin{table}[htb] 
\centering
\caption{The flow of the executing {\sf Algorithm 2} for ${\rm P}_*(f)$ of Figure \ref{fig:algorith}}
\begin{tabular}{|c|c|c|c|c|c|} 
\hline
traverse path & Steps & \underline{arc selected} in $A^{\hat x}_R$   & ${\hat x}$ &
 ${\cal H}(x)$ &  output  \\
\hline
 $x^*$ & Initializing & &  $x^*$ & ${\cal H}(x^*)$ & $x^*$ \\
 &  RS  & $[\underline{-2(2)}, -1-2 $] & & &  \\
$x^* \to  x_1 $ & RT &$-2(2)$&  $x_1$ & ${\cal H}(x_1)$ &$x_1$ \\
 &  RS &$[\underline{-1-2}]$& & & \\ 
$x_1 \to x_2 $ &  RT &$[{-1-2}]$&  $x_2$ &${\cal H}(x_2)$  & $x_2$ \\ 
 & RS &$ \emptyset $ & & & \\
$x_2 \to x_1$ & FT &$ [1+2] $ & $x_1$&${\cal H}(x_1)$& \\
& RS & $ [-1-2] $(searched) & & & \\
$x_1 \to x^* $& FT & $ 2(2)$  & $x^*$& ${\cal H}(x^*)$ & \\
& RE & $[-2(2),\underline{-1-2]}$ & & & \\
$x^*  \to x_3 $ & RT & $[-1-2]$ & $x_3$ & ${\cal H}(x_3)$  &  $x_3$ \\
 & RS  & $[\underline{1-2}]$ &  & & \\
$x_3 \to  x^*$&  FT  & $[1+2]$  & $x^*$& ${\cal H}(x^*)$ & \\
Stop &  &&& & \\
\hline
\end{tabular}\\
\leftline{\small We use the first letter to represent the step of {\sf Algorithm 2}, e.g., ``RS'' representing ``Reverse}\par
\leftline{\small  search'', and so on.}
\label{tab:algorithm_t}
\end{table}
We use the first letters to represent the steps of {\sf Algorithm 2}, e.g., ``RS'' representing ``Reverse search'', and so on.
\bigskip

Before demonstrating the complexity of {\sf Algorithm 1}, we present the computation of capacity functions  (\ref{eq:alg_2_g}) and (\ref{eq:alg_2_g2}). 
Capacity functions linked to the adjacent vertex can be computed efficiently without addressing the bisubmodular minimization issue. 
 
Let $x \in {\rm P}_*(f)$ be a vertex. 
From Extreme point theorem \ref{th:ext}, $x$ must lay in a base polyhedron ${\rm B}_{(S,T)}$ for some $(S,T)\in 3^N$. 
We denote $(S,N\setminus S)$ by $(S,N-S)$, and simply $\{i\}$ by $i$ in the sequel. 

Recall the signed graph ${G}'(x,\partial a)$ and its transitive closure ${\tilde G}'(x,\partial a)$ defined in Theorem \ref{th:adjac}, we have the following Lemma \ref{capacity} which presents how to efficiently compute the capacity functions. 

\begin{lemma} \label{capacity} 
Let $x, x'\in {\rm P}_*(f)$ be two adjacent vertices, and ${\tilde G}'(x,\partial a)= {\tilde G}' (x', -\partial a)$ for $ a \in {\cal H}(x)$. 
Then, we have:
\begin{eqnarray}
 && 2c(x,i,i)=f(S,N-S)+f(S\hspace{-1mm}+i,N-S-\hspace{-1mm}i)-2f(S,N-S-i),\label{eq:c-2i} \\ 
 &&2c(x,-i,-i)=f(S,N\hspace{-1mm}-\hspace{-1mm}S)\hspace{-1mm}+ \hspace{-1mm}f(S\hspace{-1mm}-\hspace{-1mm}i,N\hspace{-1mm}-\hspace{-1mm}S+\hspace{-1mm}i)-2f(S-i,N-S),\label{eq:c-2in}
\end{eqnarray}
if arc $a$ is selfloop, $i \notin S \subset N $ in (\ref{eq:c-2i}) and $i \in S \subseteq N $ in (\ref{eq:c-2in}); 
\begin{eqnarray}
 c(x,i,j) 
=  f({\rm dep}(x,+i)^+ -i,{\rm dep}(x,+i)^- )
 + f({\rm dep}(x,+i)^+,{\rm dep}(x,+i)^-  -j) \nonumber\\
  \quad  - f({\rm dep}(x,+i)^+-i,{\rm dep}(x,+i)^- -j ) 
 - f({\rm dep}(x,+i)^+,{\rm dep}(x,+i)^- ) \nonumber\\
 \quad \qquad \qquad \qquad \qquad {\it if}\; j \in {\rm dep}(x,+i)^-, \label{eq:c_ij} 
\end{eqnarray}
for arc $a$ of type $\partial a = i+j \; (i \ne j)$.

Furthermore, we have $c(x,i,j)=c(x,j,i)$, if $j \in {\rm dep}(x,+i)^-$ and $ {\rm dep}(x,+j)^- \ne \emptyset$ or vice versa.
\medskip \\
\rm
(Proof) See Appendix. 
\pend
\end{lemma}

In Lemma \ref{capacity}, the computations do not include the adjacent vertex $x'$, only specifically in $x$.    
Computing (\ref{eq:c-2i}) and (\ref{eq:c-2in}) are relatively simple, whilst (\ref{eq:c_ij}) relies on the structure of $\mathcal{H}(x)$.
For the ``Furthermore'' part, it appears to be the symmetry of notations, we will see that it is not a simple repetition. 
Refer to the discussion provided in Appendix.

Lemma \ref{capacity} does not include the cases of computing the capacity functions $c(x,-i,j)$ and $c(x,i,-j)$.
In these cases, we have both $x$ and $'x$ in the same base polyhedron ${\rm B}_{(N,\emptyset)}$, refer to paper \cite{zhan}. 

As mentioned earlier, computing general capacity functions necessitate 
solving bisubmodular function minimization.  
The first combinatorial polynomial time algorithm for the bisubmodular 
function minimization is a result of Fujishige and Iwata \cite{FI}. 

\begin{theorem} \label{th:complexity}
Assume that there is an oracle of computing bisubmodular functions.
Then, there is an implementation of the reverse search for enumerating all vertices of a bisubmodular polyhedron with time complexity 
$\mathcal{O}(n^4|V|)$ and 
space complexity $\mathcal{O}(n^2)$. 
\medskip \\
\rm
(Proof) 
We first show the time complexity. 

In Initializing, the time for computing $x^*$ is $\mathcal{O}(n)$, computing dependence functions and Hasse diagrams of Procedure I are bounded by $\mathcal{O}(n^2)$ \cite{af4}.  

We need to evaluate whether the chosen arc on the list $A_R^{\hat x}$ is determined by local search function $g$ in Reverse search, these computations are bounded by the number of arcs, i.e., ${O}(n^2)$. 

In Reverse/Forward traverse, computing $i^*,j^*$ of (\ref{eq:local_f}) is clearly bounded by O$(n^2)$. 
The time complexity of Procedure I (described in Initializing) is $\mathcal{O}(n^2)$. 
Hence, the total time complexity for this implementation is bounded by $\mathcal{O}(n^4)$. 

Furthermore, by Lemma \ref{capacity} and the discussions below it, all the capacity functions in {\sf Algorithm 2} are computed in constant time. 

Because above operations are repeated for each vertex, hence, the total time for the implementation is bounded by $\mathcal{O}(n^4|V|)$. 

For the space complexity, in all steps, we require $\mathcal{O}(n^2)$ space for the dependence functions, Hasse diagrams, and other constant space for $i$, $j$, ${\hat x}$, etc.  
Hence the total space complexity is $\mathcal{O}(n^2$).
\pend
\end{theorem}

We call an enumeration algorithm \emph{polynomial delay} if the time between two consecutive outputs is bounded by a polynomial function of the input size in the worst case. 
\begin{proposition} \label{th:delay}
Assume that there is an oracle of computing bisubmodular functions.
Then, there is an implementation {\sf Algorithm 2} for enumerating all vertices of a bisubmodular polyhedron with an $\mathcal{O}(n^6)$ delay. 
\medskip \\
\rm
(Proof)
We employ reverse search {\sf Algorithm 1} in {\sf Algorithm 2}, which includes {a spanning tree} defined by a local search function. 
 
Each vertex of the spanning tree conforms to a vertex of a bisubmodular polyhedron,
and the arc of the tree points to adjacent vertices of the bisubmodular polyhedron.
Traverses are along the tree by depth first search implicitly. 
The diameter of a bisubmodular polyhedron is bounded tightly by $n^2$ \cite{msz_2023,zhan05} which is the deepest length of the spanning tree.

By the same discussions as the proof of Theorem \ref{th:complexity}, the worst computation at each vertex is bounded by $\mathcal{O}(n^4)$.  
 
Therefore, the time of computation between two consecutive outputs (vertices), the delay complexity of {\sf Algorithm 2}, is bounded by $\mathcal{O}(n^6)$.  
\pend
\end{proposition}
\section{Concluding remarks} \label{sec:conclude}
For a bisubmodular polyhedron, the number of its vertices varies from $1$ to $2^n n!$ extensively, where $n$ is the dimension of the underline space of the polyhedron. 
Hence, efficiently enumerating all the vertices is essential and a non-trivial challenge.

We proposed {\sf Algorithm 2} for enumerating all the vertices of a bisubmodular polyhedron in $\mathcal{O}(n^4|V|)$ time and $\mathcal{O}(n^2)$ space complexity (Theorem \ref{th:complexity}), by employing Avis and Fukuda's reverse search \cite{af}. 
We also show that {\sf Algorithm 2} is $\mathcal{O}(n^6)$ delay (Proposition \ref{th:delay}). 
{\sf Algorithm 2} is based on the adjacency characterized by vertices' signed posets investigated by Ando and Fujishige \cite{af1,af4}. 

To avoid redundant operations and achieve efficiency, the following two are the main efforts.
\vspace{-1mm}
\begin{itemize}
\item[-] We deliberately define a local (reverse) search function, and also a (forward) search list to skip some traverses (and building the signed posets).
\vspace{-2mm}
\item[-] The capacity function is required in implementing {\sf Algorithm 2}.
We present a computation with constant complexity in Lemma \ref{capacity} without addressing bisubmodular minimization issues, which considerably reduced the complexity of {\sf Algorithm 2}.  
\end{itemize}
\vspace{-1mm}
\section*{Acknowledgments}
\vspace{-1mm}
\addcontentsline{toc}{chapter}{Acknowledgments}

\vspace{-2mm}
The authors are grateful to Dr. Kazutoshi Ando for giving useful comments on the original version of the present article. 

\section*{Data availability}
\vspace{-2mm}
No data was used for the research described in the article.
\vspace{-2mm}

\section*{Funding}
\vspace{-1mm}
The first and third authors' work was supported partially by JSPS KAKENHI Grant Number, 20K04973 and 20H05964, 20K04970, respectively.
\vspace{-2mm}

\appendix
\section{The proof of Lemma \ref{capacity}}
It should be pointed out that $x$ and $x'$ of Lemma \ref{capacity} representing the adjacent \emph{vertices} of ${\rm P}_*(f)$ is crucial. 
Our proof is based on the results of the paper given by Ando and Fujishige in \cite{af4}. 

It is known that all the maximal $(X,Y) \in 3^N$ in a $\{\sqcup, \sqcap\}$-closed family ${\cal F}\subseteq 3^N$ have the same set $X \cup Y$ \cite{af4}. 
Here, the \emph{maximal} means the operation $\sqsubseteq $ defined in (\ref{eq:sq_def_c}). 
We call the set $X \cup Y$ the \emph{support} of ${\cal F}$ and denote it by Supp$({\cal F})$.  

We call ${\cal F}$ \emph{spanning} if Supp$({\cal F})=N$, and call ${\cal F}$ \emph{pre-spanning} if $|{\rm Supp}({\cal F})|=|N|-1$.
A $\{\sqcup, \sqcap\}$-closed family ${\cal F}$ with ($\emptyset, \emptyset) \in {\cal F}$ is called \emph{simple} if for each distinct $i,j \in N$, there exists $(X,Y) \in {\cal F}$ that separates $i$ and $j$. 
We call ${\cal F}$ \emph{pre-simple} if for each distinct $i,j \in N$,  
except for one fixed pair of elements, separated by an element of ${\cal F}$.
 
From Extreme point theorem \ref{th:ext}, we know that the $\{\sqcup, \sqcap\}$-closed family ${\cal F}(x)$ is spanning and simple for each vertex $ x \in {\rm P}_*(f)$.    
Let ${\cal F}(x,x')$ be the tight set of the edge of ${\rm P}_*(f)$) comprising the two adjacent vertices $x$ and $x'$.
Then, ${\cal F}(x,x')$ is pre-spanning if arc $a$ in Theorem \ref{th:adjac} is a loop, and ${\cal F}(x,x')$ is pre-simple if $a$ is not a loop.
If $x \ne x'$, we have 
\begin{equation} \label{eq:f_x_x'}
 {\cal F}(x,x') \subset {\cal F} (x) \quad {\rm and} \quad {\cal F}(x,x') \subset {\cal F}(x'). \hspace{20mm} 
\end{equation}

We prove (\ref{eq:c-2i}), the equality (\ref{eq:c-2in}) can be shown similarly.

If ${\cal F}(x,x')$  is pre-spanning, the edge ${\cal F}(x,x')$ of P$_*(f)$ is defined by $n-1$ (non-redundant) equalities satisfying (\ref{eq:f_x_x'}). 
This is possible if and only if $i$ is on the \emph{top} of the ideals of ${\cal I}(x)$ and ${\cal I}(x')$, or the last element in the associated chain (\ref{eq:chain_1}) computing vertices $x$ and $x'$ (recall Extreme point theorem \ref{th:ext}, and refer to \cite{msz_2023} for the similar discussion). 
Thus, we have:
\begin{eqnarray}
&-x(i) = -(x(\{N-S\}-\{N-S-i\}))=f(S,N-S)-f(S,N-S-i). \nonumber \label{eq:x'} \\
& \quad x'(i) = x'(\{S+i\}-S)=f(S+i,N-S-i)-f(S,N-S-i). \qquad \qquad\label{eq:x} \nonumber
\end{eqnarray} 
The definition (\ref{eq:cap1}) of $c(x,i,i)$ means 
$$ 
x'=x+2c(x,i,i) \chi_i. \nonumber
$$ 
Therefore, we have 
\begin{eqnarray}
  2c(x,i,i) &=&x'(i) - x(i)  \nonumber \\
  &=&  f(S,N-S) +f(S+i,N-S-i) - 2f(S,N - S-i). \nonumber
\end{eqnarray}
This ends the proof of (\ref{eq:c-2i}). 
\smallskip 

To calculate the capacity function $c(x,i,j)$ when $(i \ne j)$, we first show the following Claim I. 
\smallskip \\
{\bf Claim I}: \emph{For an acyclic bidirected graph $G(N,A;\partial)$, the signed subset $(X,Y) \in 3^N$ is an ideal of $G(N,A;\partial)$ 
if and only if it is an ideal of its closure ${\tilde G}(N,A;\partial)$.}
\smallskip 

The if part is clear since ${ G}(N,A;\partial) \subseteq {\tilde G}(N,A;\partial)$. 
Conversely, if $\forall a, a' \in { G}(N,A;\partial)$ such that  $\langle \partial a, \chi_{(X,Y)} \rangle \leq 0 $ and  $\langle \partial a', \chi_{(X,Y)} \rangle \leq 0$ for some $(X,Y) \in 3^N$, 
then, for arc $ \partial a'' =  \partial a +  \partial a' \in {\tilde G}(N,A;\partial)$ by the definition of the transitive closure, we have     
$$ 
\langle \partial a'', \chi_{(X,Y)} \rangle =\langle \partial a, \chi_{(X,Y)} \rangle + \langle \partial a', \chi_{(X,Y)} \rangle \leq 0.  
$$ 
The loop arc can be treated similarly. 
This ends the proof of Claim I.
\medskip

We turn to the proof of (\ref{eq:c_ij}).

Recall that the set $\mathcal{F}(x,x')$ is pre-simple.
From vertex $x$, its adjacent vertex $x'$ is acquired by swapping two elements $i$ and $j$ in chain (\ref{eq:chain_1}). 
It is clear that $x(i)+x(j)=x'(i)+x'(j)$ by the computation of (\ref{eq:chain_1}) (all other components of $x$ and $x'$ remain the same).
Therefore, to compute $c(x,i,j)$, it is enough to show the following Claim II.
\medskip \\
\noindent
{\bf Claim II}: From the assumption of Lemma \ref{capacity} with $i \ne j$, we have 
\begin{eqnarray} \nonumber
 &&  x(i) =f({\rm dep}(x,+i)^+,{\rm dep}(x,+i)^-  )-  f({\rm dep} (x,+i)^+ -i,{\rm dep}(x,+i)^-  ), \qquad \label{eqn:ipj} \\
 && x'(i)= f({\rm dep}(x,+i)^+,{\rm dep}(x,+i)^- -j )-
 f({\rm dep}(x,+i)^+ -i,{\rm dep}(x,+i)^- -j).\label{eqn:ipj2} \nonumber
\end{eqnarray} 

Suppose $j \in {\rm dep}(x,+i)^-$ in (2b) of Subsection \ref{ssec:signed}. 
What we need to show is: 
\smallskip \\
\noindent
(C2-1) $({\rm dep}(x,+i)^+ , {\rm dep}(x,+i)^- ) \in {\cal F}(x)$, \\ 
(C2-2) $({\rm dep}(x,+i)^+ -i, {\rm dep}(x,+i)^- ) \in {\cal F}(x)$, \\ 
(C2-3) $({\rm dep}(x,+i)^+ -i, {\rm dep}(x,+i)^- -j) \in {\cal F}(x')$, \\
(C2-4) $({\rm dep}(x,+i)^+ , {\rm dep}(x,+i)^- -j) \in {\cal F}(x')$.   
\smallskip 

The case of (C2-1) is obvious from the definition of the dependent function. 

Case (C2-2) is immediate since $i$ is the maximal element of ${\rm dep}(x,+i)$ in the lower ideal $\mathcal{I}(x)$, and ${\cal F}(x)$ is spanning. 

We show (C2-3).  
By the definition of dep$(x,+i)$, we have $i \in {\rm dep}(x,+i)^+$. 
Since $x'$ is the adjacent vertex of $x$, from Theorem \ref{th:adjac} we have an arc $-i-j$ in ${\cal H}(x')$ and $i \in {\rm dep}(x',-j)^+$. 
We have $j$ being  the maximal element of dep$(x',-j)$ in lower ideal $\mathcal{I}(x')$ similar to (C2-2).
Then, we obtain (C2-3) from the pre-simple property of ${\cal F}(x,x')$ and (\ref{eq:f_x_x'}).  

Case (C2-4) is shown as follows. 

For each $(X,Y) \in {\cal F}(x)$ with $i \in X$ and $j \in Y$, i.e., $(X,Y)$ is an ideal of $G(x)$, it is clearly an ideal of $G' (x,-i-j)$.  
We know that $(X,Y)$ is also an ideal of $ {\tilde G'}(x, -i-j)$ from Claim I.
Then, $(X,Y)$ is also an ideal of $G(x')$ since $ {\tilde G' }(x',i+j)={\tilde G'} (x, -i-j)$, which means $(X,Y) \in {\cal F}(x')$. 

Therefore, we have ${\rm dep}(x,+i) \in {\cal F}(x')$ since 
the assumption of $j \in {\rm dep}(x,+i)^- $, ${\rm dep}(x,+i)\in {\cal F}(x)$, and the discussions above.
Now, we can show (C2-4) by the similar discussions of (C2-3).  
There is no arc of the type $\partial a =i+j $ in ${G}'(x', -i-j)$ and in $G(x')$ from the acyclic property of $G(x') $. 
The only arc between $i$ and $j$ in $G(x') $ is $ \partial a' = -i -j $ by Theorem \ref{th:adjac}. 
Thus, $({\rm dep}(x,+i)^+ , {\rm dep}(x,+i)^- -j)$ is an ideal of $G(x')$ from the definition of ideal. 

The case of $i \in {\rm dep}(x,+j)^-$ is treated similarly. 
This finishes the proof of Claim II. 
\medskip

The capacity function $c(x,i,j)$ is obtained from the result of Claim II by following arithmetic calculation:
\begin{eqnarray}
  c(x,i,j) &=& x'(i) - x(i)  \nonumber \\
  &=& \left(f({\rm dep}(x,+i)^+,{\rm dep}(x,+i)^-  -j)
   -f({\rm dep}(x,+i)^+-i,{\rm dep}(x,+i)^- -j )\right)\nonumber \\
  & & - \left(f({\rm dep}(x,+i)^+,{\rm dep}(x,+i)^-)
    - f({\rm dep}(x,+i)^+ -i,{\rm dep}(x,+i)^-)\right)\nonumber \\
  &=& f({\rm dep}(x,+i)^+ -i,{\rm dep}(x,+i)^- )
     + f({\rm dep}(x,+i)^+,{\rm dep}(x,+i)^-  -j)\nonumber \\
  & & - f({\rm dep}(x,+i)^+-i,{\rm dep}(x,+i)^- \hspace{-1mm}-j ) 
      - \hspace{-1mm}f({\rm dep}(x,+i)^+,{\rm dep}(x,+i)^- ), 
\end{eqnarray}
This ends the proof of (\ref{eq:c_ij}). 

Finally, we show the ``Furthermore" part. 

For arc of type $i+j$ in ${\cal H}(x)$, either $j\in {\rm dep}(x,+i)^-$ or $i \in {\rm dep}(x,+j)^-$ holds, from Claim II we can verify a relation by
\begin{equation} \begin{array}{ll}
& f({\rm dep}(x,+i)^+,{\rm dep}(x,+i)^- -j )-
  f({\rm dep}(x,+i)^+ -i,{\rm dep}(x,+i)^- -j ) \\
 & + f({\rm dep}(x+j)^+,{\rm dep}(x,+j)^- )- 
  f({\rm dep}(x,+j)^+,{\rm dep}(x,+j)^-  -i) \\
 = & f(({\rm dep}(x,+i)^+,{\rm dep}(x,+i)^- -j) \sqcup ({\rm dep}(x,+j)^+,{\rm dep}(x,+j)^-)) \\  
 & +f(({\rm dep}(x,+i)^+,{\rm dep}(x,+i)^- -j) \sqcap ({\rm dep}(x,+j)^+,{\rm dep}(x,+j)^-)) \\
 &-f(({\rm dep}(x,+i)^+-i,{\rm dep}(x,+i)^--j) \sqcup ({\rm dep}(x,+j)^+,{\rm dep}(x,+j)^--i)) \\ 
 &-f(({\rm dep}(x,+i)^+-i,{\rm dep}(x,+i)^--j) \sqcap ({\rm dep}(x,+j)^+,{\rm dep}(x,+j)^--i)) \\
 = &0,
\end{array} \end{equation}
where the first equality follows from the fact that ${\cal F}(x)$ is closed under $\sqcap$ and $\sqcup$, 
the second one follows from $ i \in {\rm dep}(x,+j)^-$, 
$i \notin {\rm dep}(x,+j)^+$ and $i \notin {\rm dep}(x,+i)^-$. 
The result indicates that $x'(i)$ computed by the assumption $j \in {\rm dep}(x,+i)^-$ in (\ref{eq:c_ij}) is equal to $x'(i)$ computed by $i \in {\rm dep}(x,+j)^-$. 
Furthermore, if $j \in {\rm dep}(x,+i)^-$ and ${\rm dep}(x,+j)^- \ne 
\emptyset$, we always have  $i \in {\rm dep}(x,+j)^-$, a relation mentioned in \cite{af1}. 
Therefore, either $j \in {\rm dep}(x,+i)^-$ or $i \in {\rm dep}(x,+j)^-$, they are equivalent.  

This ends the proof of Lemma \ref{capacity}. 
\pend




\end{document}